# On propagation of positive and negative streamers in air in uniform electric fields

G V Naidis and N Yu Babaeva

Joint Institute for High Temperatures, Russian Academy of Sciences, Moscow 125412, Russia

E-mail: gnaidis@mail.ru

**Abstract** Recently published results of numerical simulations of positive and negative streamers propagating in uniform electric fields in air are analyzed here in the framework of an analytical approach. Obtained approximate relations between the streamer radius, velocity and length, depending on the value of applied electric field, are in reasonable agreement with the results of numerical simulations.

Ionization waves – streamers formed in dielectric media at application of sufficiently strong electric fields are actively studied as key elements in pre-breakdown phenomena [1,2]. Recently a number of papers have been published [3-10] on numerical modeling of streamers propagating in air in uniform electric fields. Various aspects of the dynamics of accelerating and decelerating positive and negative streamers have been studied in a wide range of external conditions: gas pressure, gap length, applied voltage, etc. In this work, results obtained in [3-10] and in earlier publications on this topic [11,12] are analyzed in the framework of approximate analytical approach [13,14].

The dynamics of streamers developing in short regions of strong electric field, e.g. near needle protrusions at plate electrodes, and later propagating in much longer regions of relatively weak uniform electric field, e.g. between plate electrodes, is considered. The character of streamer motion in uniform fields depends on the relation between the values of the applied field $E$ and the field in the streamer channel $E_c$. Below the conditions are considered when the losses of electrons in the channel, due to attachment to gas molecules and electron-ion recombination, during streamer propagation are small. In this case, variation of the field $E_c$ along the channel is rather weak. The fall $U_c$ of electric field potential along the head and channel of a streamer having the length $L$ can be evaluated as $U_c = U_h + LE_c$, where $U_h$ is the potential of streamer head.

On the other hand, the potential of applied field at the distance $L$ from the point of streamer initiation is equal to $U_0 + LE$, where $U_0$ is the fall of potential in the small region near initiating electrode. According to these relations, the head potential $U_h = U_0 + L(E-E_c)$ varies during streamer propagation proportionally to the streamer length, increasing at $E > E_c$ and decreasing at $E < E_c$. Stable propagation of streamers, with the velocity and radius independent of the streamer length, takes place at the applied field value $E_s$ equal to the field in the channel $E_c$.

The streamer head potential $U_h$ is nearly proportional to the product of streamer radius $R$ and electric field $E_h$ in the streamer head. Results of numerous experimental and computational studies show that in conditions of streamer propagation in the applied fields $E$ not much stronger than $E_s$ the values of $E_h$ vary in a rather narrow interval. In particular, for streamers in air the reduced electric field in the head $E_h/\delta$, where $\delta$ is the ratio of gas density to its normal value (at room temperature and atmospheric pressure), is typically within the intervals 120-160 kV/cm and 100-120 kV/cm for positive and negative streamers, respectively [14]. Therefore, the streamer radius $R$ varies with the length $L$, similarly to $U_h$, nearly proportionally to $L(E-E_s)$, according to the equation [13]

$$dR/dL = a(E/E_s\text{-}1). \quad (1)$$

The reduced streamer radius $R\delta$ is related with the streamer velocity $V$ nearly as $V = bR\delta$, where the coefficient $b$ is governed by the value $E_h/\delta$ of reduced electric field in the streamer head [13,14]. The use of this relation, together with equation (1), gives, in assumption that variation of $E_h/\delta$ and hence of the coefficient $b$ during streamer propagation is weak (note that this assumption is not valid for decelerating streamers approaching stopping points, see below), the equation

$$dV/d(L\delta) = bdR/dL = ab(E/E_s\text{-}1). \quad (2)$$

The coefficient $a$ in equation (1) is roughly estimated, using the relations presented above, as $a \sim E_s/E_h$. Note that accurate values of the parameters $E_s$, $a$ and $b$ in equations (1) and (2) cannot be obtained in the framework of approximate analytical approach. Below they are estimated using the results of computations of positive [3-9,11] and negative [10,12] air streamers propagating in uniform applied fields. The profiles of streamer radius and velocity versus the length obtained in these computations are rather close to linear (at least as some rather long intervals of the streamer length variation), with slopes depending on the value of applied field.

Note that the data on streamer radius presented in the mentioned papers refer to the optical radius [4,7,10] or to the streamer head radius [3,6,8], the latter corresponding to the radial coordinate, just behind the streamer head, corresponding to the maximum of radial electric field. Though generally these radii do not coincide, in the case when the streamer head has a hemispherical shape the head radius is close to the optical radius [15].

In figure 1 the values $dR/dL$ estimated for positive streamers using the plots of the streamer radius versus the length, presented in [3,4,6,8] for pressure $P$ = 1 bar and in [7] for $P$ = 0.1 bar, are shown versus the reduced applied electric field $E/\delta$. Linear dependence (1), with parameters $a$ = 0.011 and $E_s/\delta$ = 4.7 kV/cm, is also given. It is seen that all the data are close to the same linear dependence. Note that at given $E/\delta$ the slopes $dR/dL$ evaluated using the data obtained in [3] for the head radius and, for similar conditions, in [4] for the optical radius are close to each other, hence it can be expected that in these conditions the relative difference between the head and optical radii is rather small.

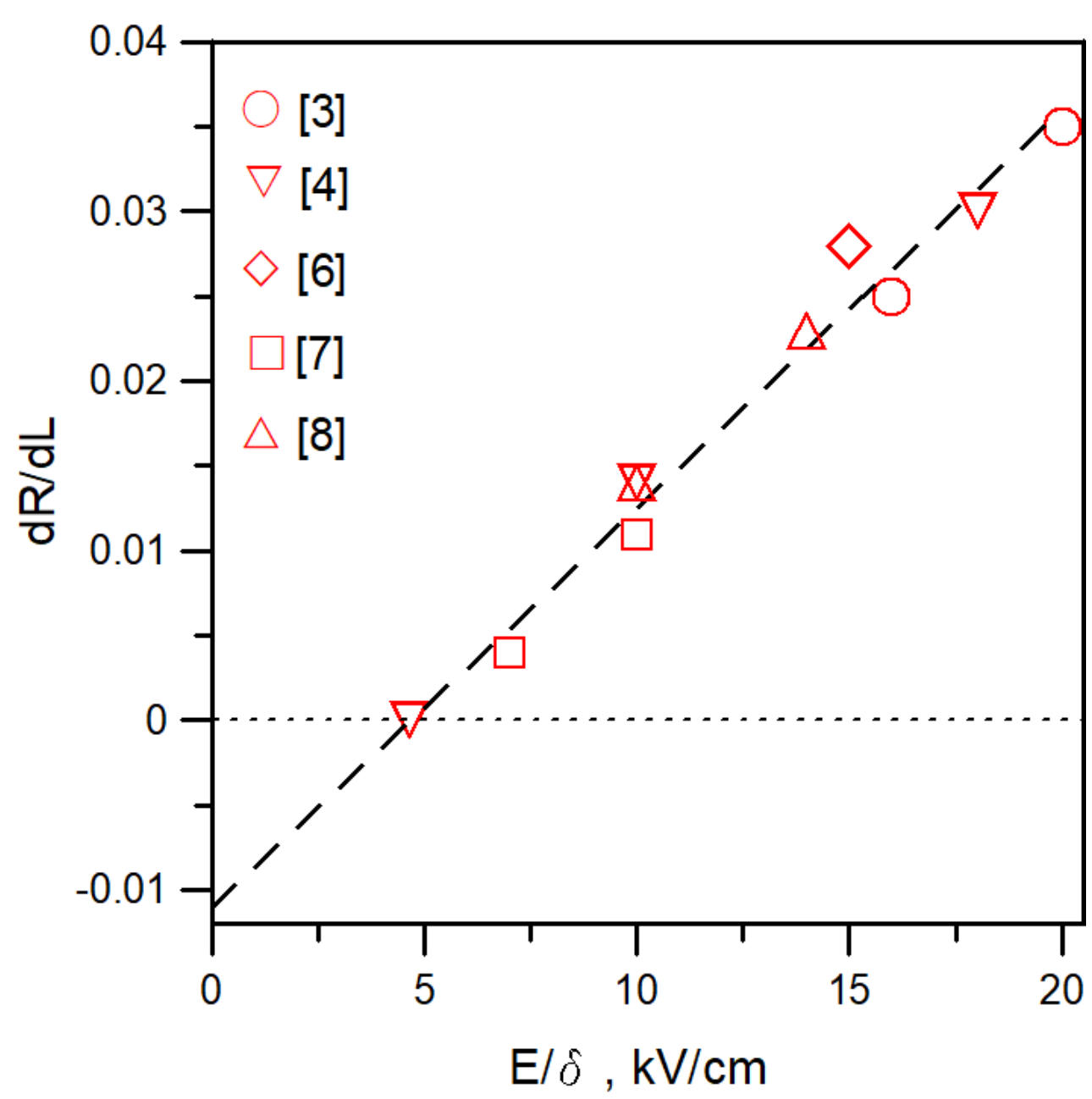

Figure 1. The rate of variation of positive streamer radius with length versus reduced electric field. Points – estimates obtained using calculated data for pressure 1 bar [3,4,6,8] and 0.1 bar [7]. Dashed line – equation (1) with parameters $a$ = 0.011 and $E_s/\delta$ = 4.7 kV/cm.

In figure 2 the slopes $dV/d(L\delta)$, evaluated for positive streamers using the profiles of streamer velocity versus the length presented in [3,4,6,7,9,11], are shown versus the reduced applied electric field $E/\delta$. The linear dependencies (2) are also given, corresponding to three values of the parameter $b = V/(R\delta)$: 4.2, 2.6 and 1.5 $ns^{-1}$, estimated in [14] for positive streamers at $E_h/\delta$ = 160, 140 and 120 kV/cm, respectively. It is seen that all the data are within the region between lines 1 and 3, corresponding to the upper and lower boundaries of the interval of $E_h/\delta$ typical for positive streamers in air. The values of $E_h/\delta$ obtained in simulations of weakly accelerating streamers, about 160 kV/cm in [11] and 120 kV/cm in [4], correspond to the maximal and minimal slopes. Note that though these slopes differ substantially, the values $E_s/\delta$ of reduced electric field corresponding to zero slopes are close to each other.

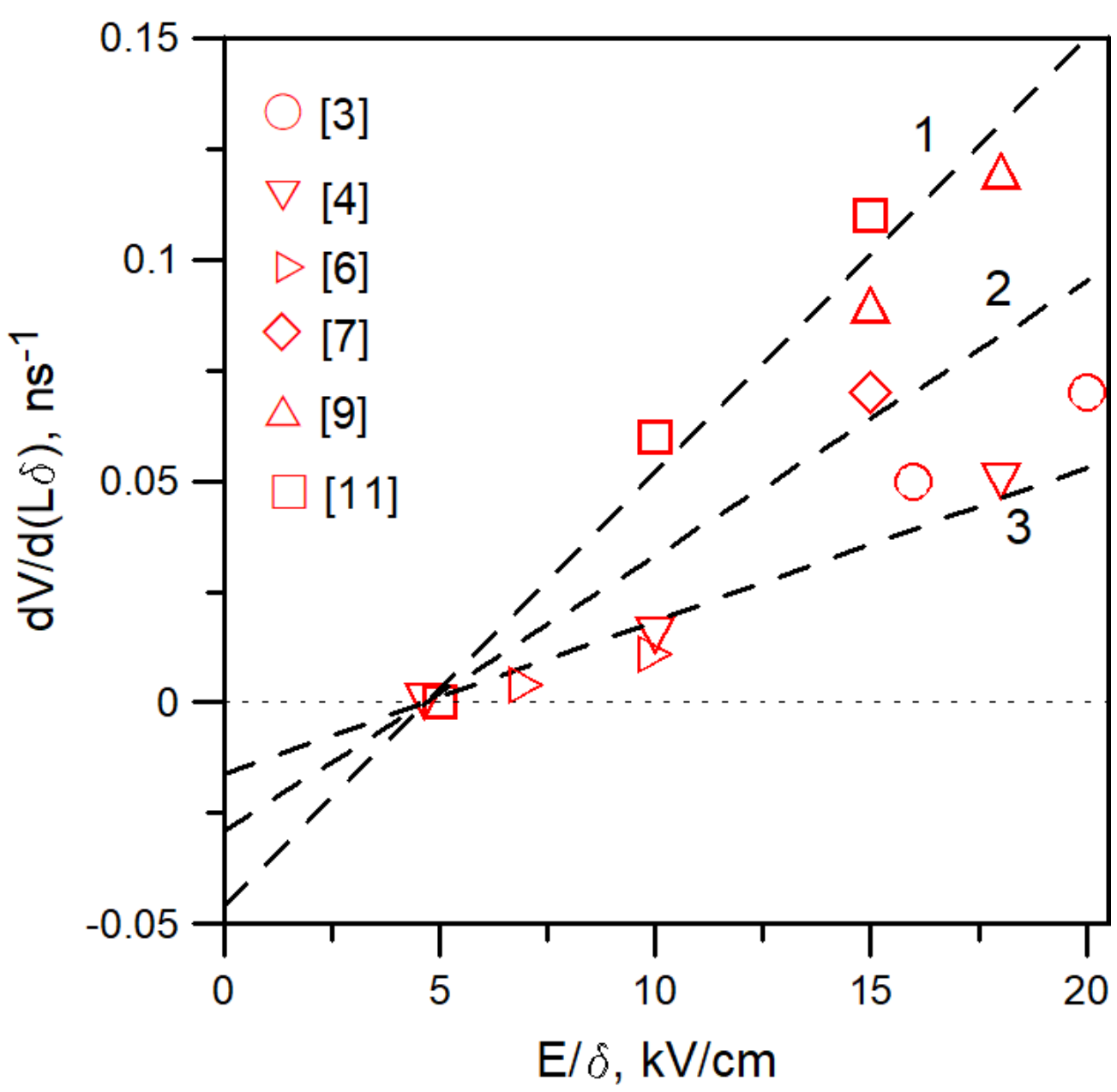

Figure 2. The rate of variation of positive streamer velocity with reduced length versus reduced electric field. Points – estimates obtained using calculated data for pressure 1 bar [3,4,6,9,11] and 0.1 bar [7]. Dashed lines – equation (2) with parameters $E_s/\delta$ = 4.7 kV/cm, $a$ = 0.011 and $b$ = 4.2 $ns^{-1}$ (line 1), 2.6 $ns^{-1}$ (line 2), 1.5 $ns^{-1}$ (line 3).

Computational data obtained for negative streamers in air can be also analyzed using equations (1) and (2), with parameters $E_s$, $a$ and $b$ differing from those for positive streamers. In figure 3 the slope $dR/dL$ evaluated using the plot of the streamer radius versus the length presented in [10] is shown, together with linear dependence (1) with parameters $a$ = 0.08 and $E_s/\delta$ = 11.7 kV/cm. Thus estimated parameter $a$ for negative streamers is about 7 times larger than that obtained above for positive streamers. This difference is partly caused by the difference, of about 3 times, between the values of $E_s/E_h$ for positive and negative streamers. Unfortunately, there is no available computational data allowing to check if the slope $dR/dL$ for negative streamers is, like that shown in figure 1 for positive streamers, rather insensitive to variation of gas pressure, discharge gap, etc.

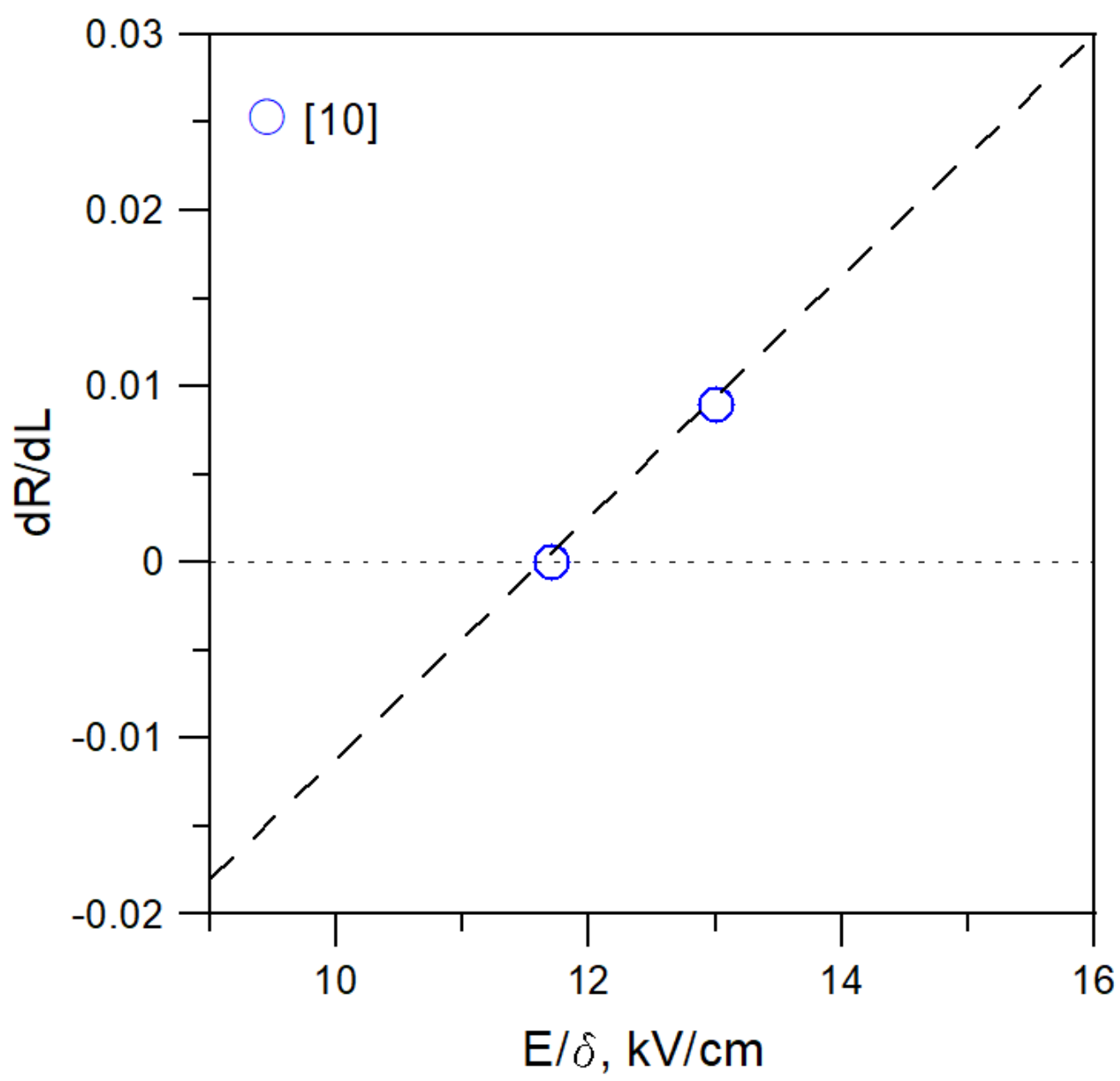

Figure 3. The rate of variation of negative streamer radius with length versus reduced electric field. Points – estimates obtained using calculated data [10] for pressure 1 bar. Dashed line – equation (1) with parameters $a$ = 0.08 and $E_s/\delta$ = 11.7 kV/cm.

Figure 4 shows the values of $dV/d(L\delta)$ versus $E/\delta$ for negative streamers, estimated using the plots of streamer velocity versus the length presented in [10,13]. The linear dependence (2) with parameters $E_s/\delta$ = 11.7 kV/cm, $a$ = 0.08 and $b$ = 1.8 ns$^{-1}$ is also given. This estimate of the parameter $b$ is close to the ratio $V/(R\delta)$ evaluated in [14] for negative streamers at $E_h/\delta$ = 110 kV/cm.

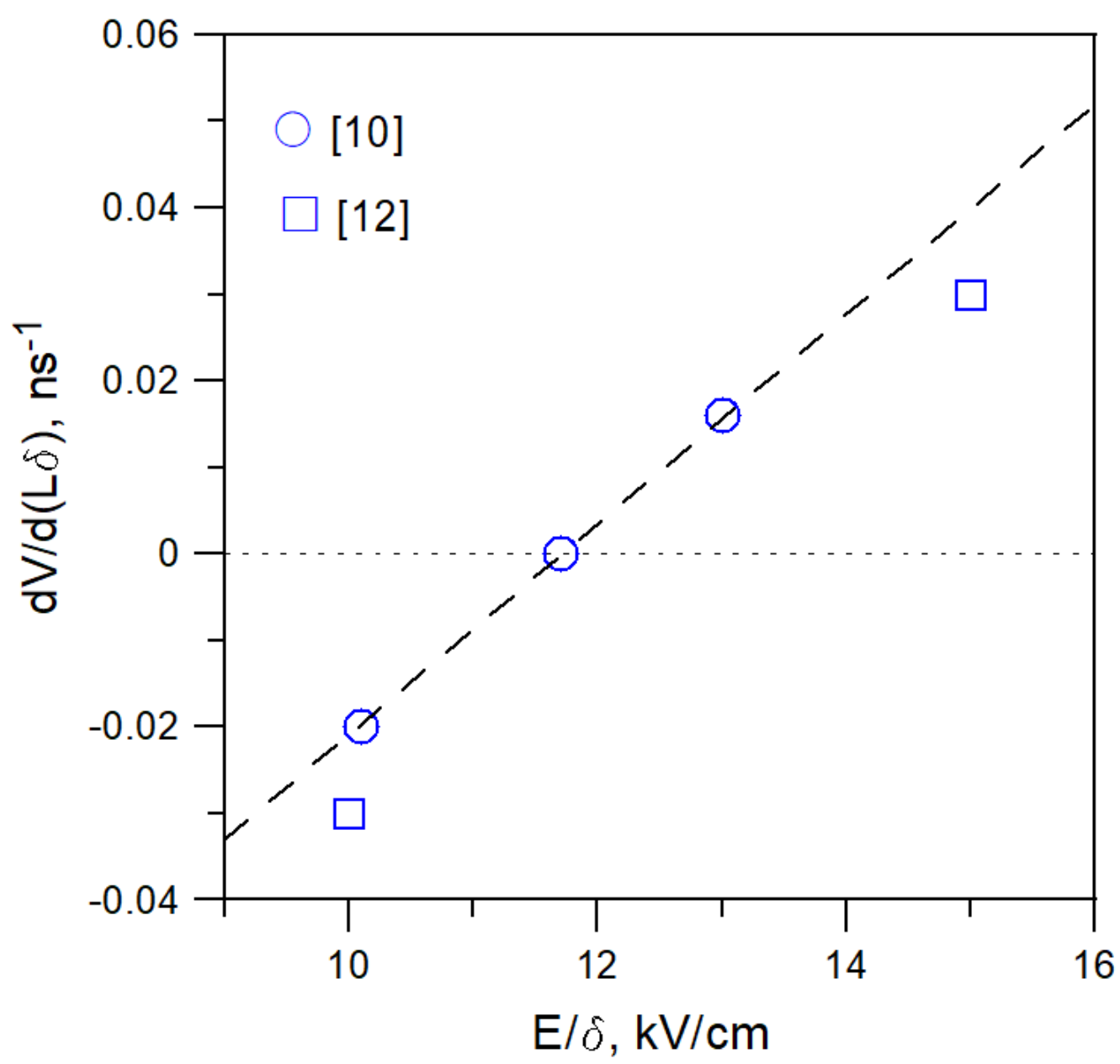

Figure 4. The rate of variation of negative streamer velocity with reduced length versus reduced electric field. Points – estimates obtained using calculated data [10,12] for pressure 1 bar. Dashed line – equation (2) with parameters $E_s/\delta$ = 11.7 kV/cm, $a$ = 0.08 and $b$ = 1.8 ns$^{-1}$.

Results presented above show that equations (1) and (2) describe reasonably well the behavior of velocity and radius of accelerating streamers. For decelerating streamers, the behavior of streamer characteristics is more complex. When streamers approach stopping points, the reduced field in the streamer head $E_h/\delta$ changes substantially, decreasing in negative streamers [10,16] and increasing in positive streamers [5,16,17].

It should be noted that obtained estimates for the parameters in equations (1) and (2) are rather approximate. Depending on external conditions, these parameters can vary in some ranges. E.g., the values of $E_s/\delta$ corresponding to steady propagation of positive streamers obtained in computations [5] vary inside the interval 4.0-5.4 kV/cm. Nevertheless equations (1) and (2), although imprecise, could be useful to easily evaluate approximately the rates of variation of the streamer velocity and radius with the length depending on the value of applied uniform field.